# Supporting Crowd-Powered Science in Economics: FRACTI, a Conceptual Framework for Large-Scale Collaboration and Transparent Investigation in Financial Markets

Jorge M. Faleiro Jr. #1, Edward P. K. Tsang #2

Working Paper WP076-15

# Centre of Computational Finance and Economic Agents, University of Essex
Wivenhoe Park, Colchester, CO4 3SQ, UK
1 faleij@essex.ac.uk
2 edward@essex.ac.uk

*Abstract*—Modern investigation in economics and in other sciences requires the ability to store, share, and replicate results and methods of experiments that are often multidisciplinary and yield a massive amount of data. Given the increasing complexity and growing interaction across diverse bodies of knowledge it is becoming imperative to define a platform to properly support collaborative research and track origin, accuracy and use of data.

This is the first paper in the pathway of an overall research that aims to define a conceptual framework to support large-scale collaboration through a streamlined computational representation and allow unquestionable transparency in the way raw data and corresponding results are obtained, modelled, used and calculated. We are calling this framework of concepts FRACTI.

The specific intent of this present paper is to introduce a *blueprint* for the computational representation of FRACTI, outlining assumptions, definitions and a review of challenges specific to the field of computational finance. This blueprint is based on two fundamental assumptions:

First, the *scientific method* is a simple and proven tool to support *transparent investigation and collaboration*, and should be used as a foundation for investigative procedures in the field of computational finance and in an ever growing set of correlated disciplines, even outside of academia – in corporate and regulatory arenas.

Second, there is a *computational model* to represent shareable components called *contributions* and support large-scale investigation in finance across different asset classes, correlated disciplines and in different patterns of storage and frequency.

This paper starts by defining a set of *methods* leveraging scientific principles and advocating the importance of those methods in multidisciplinary, computer intensive fields like computational finance.

The next part of this paper defines a class of systems called *scientific support systems*, vis-à-vis usages in other research fields such as bioinformatics, physics and engineering. We outline a basic set of fundamental concepts, and list our goals and motivation for leveraging such systems to enable large-scale investigation, "crowd powered science", in economics.

The core of this paper provides an outline of FRACTI in five steps. First we present definitions related to scientific support systems intrinsic to finance and describe common characteristics of financial use cases. The second step concentrates on *what* can be exchanged through the definition of shareable entities called *contributions*. The third step is the description of a classification system for *building blocks* of the conceptual framework, called *facets*. The fourth step introduces the meta-model that will enable provenance tracking and representation of data fragments and simulation.

Finally we describe intended cases of use to highlight main strengths of FRACTI: application of the scientific method for investigation in computational finance, large-scale collaboration and simulation.

I. SCIENTIFIC LEARNING AND ECONOMICS

If finance were to be studied rigorously like other sciences, being able to store, share, and replicate results and methods of experiments is critical. As complexity of financial use cases increase exponentially over time and results of everyday experiments grow into massive datasets, it is becoming imperative to streamline the representation of models and to achieve unquestionable transparency in the way raw data is obtained and stored, and corresponding results are modelled, used and calculated.

As more and more individuals, in different roles and at different levels of technical understanding are required to interact, it is important to identify, share and track the provenance [1] of artifacts [2] in a way that facilitates true collaboration and allows society as a whole to benefit from improved transparency and fairness in the market.

This paper is the first publication in the pathway of our overall research that aims to contribute a conceptual framework for a simple, straightforward computational representation that can be used by individuals in a broad

---

[1] Chronology of the ownership, custody or location of historical entities [47]
[2] Observations in a scientific investigation or experiment that is not naturally present but occurs as a result of the preparative or investigative procedure [47]

financial community, in diverse roles, to communicate and collaborate transparently.

This research advocates for a process of large-scale investigation leveraging primarily open collaboration and well-known procedures dictated by the modern scientific method [1], a "crowd science" [2] organization to solve practical problems in economics and financial markets.

This conceptual Framework for Collaboration and Transparent Investigation – FRACTI – defines a set of ideas and relationships for a collaboration platform and will serve as a baseline for extensions and improvements for future research. Implementations of the ideas proposed in this conceptual framework are not within the scope of this research but are strongly encouraged.

The specific intent of this paper is to introduce a blueprint for this computational representation, outlining definitions and a review of specific challenges for the application of this class of platform in the field of computational finance [3] [4], considering the three main peculiarities of the field.

First, computational finance deals with a very unique subject of study - a shared, intertwined, complex market – that cannot be rewound. Time like life moves towards one direction. Given the usually large number of inputs to such a complex system and the apparent independency between these input variables, once an event occurs we cannot derive different futures from what the present currently describes.

Second, when taken from a recent historical perspective, computational finance has been associated with compartmented classical fields like economics, statistics and computer sciences. Most of the assumptions in classical and theoretical sciences are inherently oversimplified and flawed when trying to predict or understand the behaviour of a systemic market [5][3].

Third, modern computational finance in essence is a multidisciplinary subject. Efforts to understand the market in terms of its most fundamental structures that would allow the inference of predictability tend to rely on rather orthogonal fields of study like neuroeconomics [6], behavioural sciences [7] and analysis of market micro-events [8], among others. The interdependency of computational finance to bioengineering, neurosciences, social sciences, psychology, data, computer sciences and other fields is diffuse and difficult to correlate.

We understand that these peculiarities of our field of study - systemic complexity, lack of proper theoretical models and novelty of correlated fields of study – make research in computational finance strictly dependent on high performance computers, implementing simulation-based techniques, similar to what is used in other hard-sciences, such as physics, engineering and biophysics [9] [10].

Unfortunately, this dependency on high performance computing has driven research in computational finance to favour specialized techniques for storage and processing speed. The field has been shaped so the sheer generation of data and obscure ways to represent computational procedures is prioritized over proper control.

---
[3] While we consider it important to highlight this peculiarity, evaluating reasons for such limitations, or trying to refute or confirm them is beyond the scope of this paper.

The consequences of the absence of controls in investigation in general are well documented and it is beyond the scope of this paper to list them in details. From the lack of metrics to properly indicate and predict systemic and recurring crisis, [11] to the frequency in which fraudulent, biased or poor research has been used as a basis for the definition of public policy with catastrophic social impact [12] [13] [14] [15] [16]. Proper control procedures have to become a priority for investigation and regulation in computational finance, other sciences, and even outside of academia in public and private institutions.

One of the cornerstones of this research is the realization that computing power is required to generate and store massive amounts of data and perform intensive computations, but if computing power is important in this scale, computational controls are indispensable.

In essence, this evidence shows that the limitation factor for scientific advancement has clearly shifted from availability of storage and computational resources to proper control of investigative procedures. Computing resources should be leveraged to also provide proper controls, transparency and consistent collaboration.

Finally, this research advocates that control methods should leverage principles of the modern scientific method. These principles are proven, simple and accepted enough to be used as one of the foundations for control in methods of investigation in computational finance. In modern scientific method "each principle helps to increase the reliability and accuracy of knowledge resulting from scientific research" [1]

1. The goal of scientific investigation should be to gain objective knowledge.

2. Scientific knowledge is obtained through tests, experiments and observations. Tentative assumptions about a certain phenomenon may, however, be deduced from preexisting knowledge.

3. A hypothesis must be verifiable by some experimental or observational method.

4. Experiments must be reproducible and must have controls

5. The integrity of the data must be properly safeguarded.

These principles naturally address control requirements described previously. Drafting a parallel between those requirements and these principles, we propose a class of software conceived around the following drivers:

- Providing an easy-to-use environment for individuals themselves to create and test their own hypotheses (models)

- Providing interactive tools for individuals enabling them to execute their hypotheses and tests (scenarios) and view their results in real-time

- Simplifying the process of sharing and reusing contributions (and results) among individuals.

- Enabling individuals to track the provenance (chronology of the ownership, custody or location of

historical entities) of results and mechanics to reproduce those results.

We are calling the class of software supporting these drivers a **scientific support system**.

## II. SCIENTIFIC SUPPORT SYSTEMS

A scientific support system is a specialized form of a workflow and data management system designed specifically to compose and execute a series of computational or data manipulation steps compatible with the scientific principles.

Use of specialized systems to support a scientific method is not common in most domains of investigation, exception granted to computational biology [17] [18].

Some sources refer to this specific class of systems as scientific workflow systems. To avoid confusion with systems dedicated to generic workflow management [19] we plan on using a specific denomination of "**scientific support system**" in the scope of this research [20].

### A. Motivation

The sheer volume of data and complexity involved in financial investigation on items that should ultimately be materialized as policies or business procedures, and the rate in which demands for more data and computing power are expected to increase [21] has made computing resources indispensable, but even more so, computational controls.

We are going through times in which science has worked miracles in human advancement despite of abundant evidence of misuse of computational procedures in investigation. The unbelievable rate of cutting edge scientific experiments published in prominent journals that cannot be reproduced combined with the realization that this rate is expected to grow as complexity increases, [22] and the number of flawed experiments stored in plain excel spreadsheets that have driven the definition of broken financial policies [13] [23] [24] are only a few examples of hard consequences of the lack of proper control.

Features related to control have to move forward from the back seat on finance investigation through **control methods** associated to **accessibility**, **reproducibility**, **communication** and **collaboration**.

A computational representation must be **accessible**. The widespread use of high performance computer systems demands computer literacy from scientists, what is not always possible. There is a need for an accessible computational representation that shields the inherent complexities of modern computer systems from collaborators on any scientific field, focusing on simplicity, and therefore not requiring specialized computer literacy.

An abundance of computational power requires a potentially obfuscated representation of ways to transform information, yielding massive amounts of data. The paradoxical condition in which modern investigative procedures are entangled require more computing power, which enable to transform more data, and as a consequence a higher risk of uncontrolled models and massive amounts of untraceable data, that will in turn require more elaborate techniques and computing resources to trace and decipher that data, the "informatics crisis" [17]. Without proper controls around provenance, versioning of data and models, achieving proper **reproducibility** of scientific procedures is a challenging task to accomplish.

The massive amounts of computational inputs and outputs, as well as ways to transform the latter into the former have to be properly represented. Ways to **communicate** results through proper visualization and computational representation is crucial. The massive amount of data generated as input and output cannot be represented to humans the same way as they are to computers [25]. In order to make research truly useful we need human-friendly ways to visualize evidences. At the same time, communicating methods and procedures, regarded as of greater importance than explanatory texts and figures as experimental outputs [26] [27] cannot be addressed differently than other items that require human visualization.

The final objective of an environment that provides accessibility, reproducibility, and communication is **collaboration**. Collaboration is allowing results from one experiment to be seamlessly utilized by other experiments, allowing extensions on models and data to fit additional scenarios, at the same time tracking ownership of each revision or improvement. Collaboration can occur only by exchange of artifacts that are traceable. Ways in which artifacts where produced and utilized have to be transparent to the overall community.

### B. Principles

Principles of scientific support systems must parallel those of the scientific method described in Section [I] in terms of definition and testing of a hypothesis, as well as transparently tracing and safeguarding of the underlying data as scientific evidence, specifically:

- Allow the definition of an a theoretically driven hypothesis
- Allow an hypothesis to be tested
- Allow an hypothesis to be reproduced and verified by independent parties
- Allow assumptions about a hypothesis to be deduced from historical data
- Safeguard historical data

Given the fact that these same scientific principles are used in various branches of science, after review and adaptation this scientific support framework can be leveraged for the same features in other research fields.

## III. FRACTI

The objective of a conceptual Framework for Collaboration and Transparent Investigation – FRACTI – is by now established and clear: to define a platform for a scientific support system that is focused on **transparent collaboration**, **repeatability of results**, **accessibility** and **openness**.

**Transparent Collaboration**: Support for transparent definition of large datasets following a common representation and visualization. Shared items can be examined in details and re-executed against different scenarios by different groups of users. Collaborators can easily **define and back test their own hypotheses**, support **sharing, tracking and provenance** of contributions, ensure that results are replicable and in this sense, playing a role of a **scientific support system** [20] [17].

**Reproducibility**: Scientific approach to analytical research, a scientific requirement: models and scenarios have to be reproducible by anyone. Large sets of data and models can be re-executed, allowing different organizations and individuals to easily replicate results

**Accessibility**: End users do not have to be proficient in computer science in order to be able to use, collaborate or visualize models or scenarios in the framework

**Openness**: Instances of the meta-model representing a specific configuration, execution or simulation can be exchanged across environments or different implementations. Data and method of an investigation can be traced regardless of ownership, origin or location of a contribution.

The conceptual layout of this framework is in essence a proposal to **represent knowledge** in the specialized field of economics through abstractions called **models**.

Despite of a long history of academic work attempting similar tasks in a variety of domains [28] [29], most works have been concentrating on comparative analysis or evaluating properties of specific representations. This research on the other hand follows a role-based definition of knowledge representation in which a description of a knowledge system is defined in terms of five core roles a specific representation plays [30].

Firstly, models are **surrogates**. On that sense a surrogate is by definition a substitute for the target idea itself, and as such measurement of how far or how close this surrogate is from calculations it intends to represent is secondary or irrelevant.

Second, models should be able to **define human expressions**. Models should define measurements and concepts understood by humans in a language that is adequate for human consumption, even if not directly natural.

Third, models are a medium for "**pragmatic efficient computation**". Models should be able to be replicated in computers given appropriate technology and sufficient resources.

Fourth, models establish "**ontological commitments**" [31] [32] for a representation by defining "a set of decisions about how and what to see in the world". Models are approximations of a reality, and as we define them we make decisions of what to consider and what to ignore. These decisions are ontological commitments and are "not an incidental side effect but they are of essence in our representation" [30].

Lastly, models define a "fragmentary theory of **intelligent reasoning**" represented in terms of concepts and inferences, sanctioned and recommended. Models represent "some insight indicating how people reason intelligently" about a problem or investigation [30].

The **knowledge representation system** supporting these core roles is described in the following topics in terms of *what* can be shared, called in the scope of this research **contributions**, *how* to establish fundamental building blocks called **facets**, and *structural constraints* defined by FRACTI's **meta-model**.

*A. User Contributions*

In the scope of this work, the term **contribution** applies to artifacts produced by participants (users) and transferred, or contributed, to a wider community of users through a shared scientific support system.

There are special classes of **contributions**, to be detailed in future stages of this research. Contributions should tentatively cover a broad range of models, methods, and results relevant to financial sciences [23]. Some examples include datasets in small, medium or large scale; time series in low, medium or high frequency; calculation processors and visualization plots; and results related to historical and real-time execution, simulation and back testing.

Every single contribution to the framework has a set of mandatory properties in order to be defined and shared: classification, identification, record of provenance, ownership and security.

**Classification**: Contributions follow a classification system, which is under development as part of the overall research. This classification system, referred to as taxonomy of contributions, should account for core macro functions of storage, processing and visualization.

**Identification**: Contributions should be properly identified following common standards for shared identification [33] in a way to allow sharing and ownership.

**Provenance**: The platform should track chronology of the ownership, custody or location of contributions, as well as the history of associations of contributions to financial entities.

**Ownership and security**: Given the sensitive nature of contributions, the system should ensure ownership and access only when there is proper authorization and proper authentication. Under specific circumstances when allowed by policy, access by a regulatory or surveillance authority might be granted.

Contributions are **ubiquitous**. Given proper policies and authorization, access to contributions will be possible regardless of location or type of device.

The platform should be able to provide **virtually unlimited storage and computational power**. Provenance and ownership are registered against virtually infinite storage capabilities. The state of contributions is able to record all historical data and the history of associations among financial entities. The platform should allow for distributed access in order to leverage scalable computational power across multiple processors and regions.

*B. Facets*

As part of the overall research we introduced a **taxonomy of aspects, or facets**, for now are limited to four: streams, reactives, distribution and simulation.

**Streaming** aspect defines a graph-oriented Domain Specific Language to route fragments of execution meta-data bound to a generic configuration meta-data through reusable and exchangeable processors [34] [35] [36] [37] [38].

**Reactives** are an intuitive representation of primitives and formulae, in which composition of formulae from primitives and other formulae is defined declaratively [39] [40].

**Distribution** treats aspects related to scale, both in terms of computational power and storage, through an extension of the conceptual DSL to support endpoints [41].

**Simulation** supports aspects of discrete event simulation for both historical and real-time generation and re-play of time series events. This facet is bound to a simulation meta-model to represent models, modes, shocks, benchmarks and their relationships [42] [43].

Combinations of those **facets will serve as fundamental building blocks** to other more complex abstractions in the conceptual framework.

Facets define the computational representation in the framework, and details of their structure will be described in future stages of this research.

### C. Meta-Model

FRACTI's meta-model define structural constraints for associations between contributions and facts and are classified based on its use as configuration, execution or simulation meta-model.

**Configuration meta-model**: represents a versioned snapshot of a configuration of facets in time, allowing the exact definition and reproducibility of an execution flow. Instances of this meta-model will determine a sequence of execution, versions and provenance tracking of all data used to generate any specific result set.

**Execution meta-model**: represent fragments of hierarchical data that flow through one or more compatible steps of a model. Instances of an execution meta-model are related to one specific configuration meta-model.

**Simulation meta-model**: supports the registration of experiments by associations between a hypothesis and methods under verification, given by a model, each of their executions, given by shocks or modes, and final comparison of results, given by benchmarks.

### IV. CASES OF USE

The intent behind FRACTI is to provide a platform for objective application of the scientific method, large-scale collaboration across a heterogeneous community of users, and support for simulation in economics.

On this topic we present three scenarios that serve as examples of each of these cases of use, picturing a hypothetical selection of fast learning methods for neural networks [44].

### A. Application of the Scientific Method

Researcher A is studying a new method for fast learning of neural networks based on sensitivity analysis. The intent of the research is to find a fast learning method while predicting a Dow-Jones index for a given day using a historical data set for 1994-1996.

Researcher A records in the platform the following contributions:

- Hypothesis $H$ and null hypothesis $H_0$
- Pluggable implementations of six possible problem resolution scenarios: standard algorithms, linear least square, second order, adaptive step size, appropriate weights and rescaling.
- Pluggable implementation of a one-layer neural network
- Input datasets for the time series of price variations of the index, and output dataset registering learning time and fitting

Researcher A executes and collects the resulting data once for each of the 6 scenarios. For each cycle of execution and data collection, it is only necessary to switch the pluggable implementation of the resolution scenario.

Researcher A collects output datasets for each execution and contributes the results to the platform. He also records the final findings of his experiment, stating that adaptive step size is the fastest method and provides the best fitting overall.

From this point, his findings will be unquestionably and transparently bound to the method he followed (model), input data and findings, so any other participant in the community can leverage, inspect or challenge his findings.

### B. Large Scale Collaboration

Researcher B works on the same field as Researcher A. He develops a new approach called Sensitivity-Based Linear Learning Method and wants to compare the performance of that approach to earlier methods bound to findings of Researcher A.

Researcher B contributes a new pluggable implementation called SBLLM. Researcher B adds this single implementation to scenarios already contributed by Researcher A and re-runs the existing 7 scenarios.

Researcher B records the new revision of the original model, what now brings an additional scenario, and records the results of the experiment: SBLLM is now the fastest learning method and provides the best fitting. From that point on any researcher in the world can participate in this scientific search for better learning methods.

### C. Simulations

Researcher C is also studying the same subject as Researcher A and Researcher B. He believes previous findings are somewhat flawed because they failed to take into consideration a number of independent variables.

Researcher C defines a new hypothesis stating that performance of learning methods are indeed affected by at least two independent variables:

- Number of layers $N$ of the subject network;
- Generic input as a random walk of drift $D$ and variance $\sigma$

Researcher C maps each independent variable to a parameter, defining parameters $N$, $D$ and $\sigma$.

Researcher $C$ contributes a new pluggable implementation of neural network of $N$ layers, as well as a random walk generator for drift $D$ and variance $\sigma$. He also contributes a new revision of the model created by Researcher $A$, later augmented by Researcher $B$. This revised model accounts now for the random walk generator and a variable layer neural network.

Researcher $C$ executes one run of all 7 scenarios for every permutation of $N$, $D$ and $\sigma$ as well as historical data of Dow-Jones from 1994-1996 as a baseline.

Researcher $C$ contributes back to the platform his output datasets as well as an explanation of his final findings: SBLLM is a faster implementation for $N = 1$ for reasonable values of $D$ and $\sigma$. For $N > 1$, other methods were found to perform better. Researcher $C$ starts working on a research explaining possible causes.

## V. Conclusion

The objective of the overall research is to define a conceptual framework, FRACTI, to support collaboration in large scale, traceability of simulations and data, and simplified representation so that a heterogeneous user community can conduct structured, scientific investigation in computational finance.

The specific intent of this paper is to introduce a blueprint for this representation, outlining definitions and a review of specific challenges for the application of this class of platforms in the field of computational finance.

In future stages, the ongoing research aims to extend the content of this paper by adding taxonomy of contributions and detailed computational patterns, called facets, to the scope of this work. This research also aims to provide a number of showcases to exemplify concrete use of the concepts and make them publicly available to the interested community [45] [46].

The final proposition, beyond the scope of this current research, is the materialization of a computing platform based on FRACTI concepts. This platform will support the simplicity of scientific principles, leveraging modern techniques that provide virtually infinite storage and computing power, to allow transparency, control and large-scale collaboration across a heterogeneous community of users.

Such a platform will serve as a trusted environment for exchange of ideas, procedures and data related to economics in large scale. Its results can be leveraged to educate the common investor, provide reliable data for the research community, and allow proper controls for a global market surveillance that can be ultimately used for the definition of sound public policies.